# Bidirectional Controlled Quantum Teleportation Using Eight-Qubit Quantum Channel in Noisy Environments

Moein Sarvaghad-Moghaddam[1], Zeinab Ramezani[2], IS Amiri[3,4*]

[1]Quantum Design Automation Lab, Amirkabir University of Technology, Tehran, Iran

[2]Department of Electrical and Computer Engineering, Northeastern University, Boston, MA 02115, USA

[3]Computational Optics Research Group, Advanced Institute of Materials Science, Ton Duc Thang University, Ho Chi Minh City, Vietnam

[4]Faculty of Applied Sciences, Ton Duc Thang University, Ho Chi Minh City, Vietnam

Z.ramezani@northeastern.edu, *irajsadeghamiri@tdtu.edu.vn

**Abstract**

In this work, a novel protocol is proposed for bidirectional controlled quantum teleportation (BCQT) in which a quantum channel is used with the eight-qubit entangled state. Using the protocol, two users can teleport an arbitrary entangled state and a pure two-qubit state (QBS) to each other simultaneously under the permission of a third party in the role of controller. This protocol is based on the controlled-not operation, appropriate single-qubit (SIQ) UOs and SIQ measurements in the *Z* and *X*-basis. Reduction of the predictability of the controller's qubit (QB) by the eavesdropper and also, an increasing degree of freedom of controller for controlling one of the users or both are other features of this protocol. Then, the proposed protocol is investigated in two typical noisy channels include the amplitude-damping noise (ADN) and the phase-damping noise (PDN). And finally, analysis of the protocol shows that it only depends on the amplitude of the initial state and the decoherence noisy rate (DR).

**Introduction**

One of the most outstanding results of the quantum information theory in theoretical and experimental is Quantum teleportation (QT) [1]. And, the original QT protocol was proposed by Bennett et al [2]. Many theoretical and experimental papers are attributed to study of QT protocols so far [3-12]. Karlsson and Bourennane [13] firstly presented a Controlled QT (CQT) protocol, In 1998. After that, many CQT protocols with quantum channels consists of various types of entangled states (ENSs) have been introduced [14-19]. In 2013, Zha et al. [20] proposed a Bidirectional CQT (BCQT) with a five-QB cluster state. This type of protocol can transmit information of each two users simultaneously. Up to now, various BCQT protocols have been presented by using different multi-particle ESs, say five-QB EN [21], six-QB EN [22], and seven-QB EN [23].

Recently, two BCQT protocols using quantum channels composed of seven-QB ENs have been presented by Hong [24] and Sang [25]. In these schemes, Alice (A) can teleport an arbitrary two-qubit state (QBS) to Bob (B) and B can teleport an arbitrary single-QBS to A under the control of the third person (named Charlie (C)). After that, Li and Jin [26] presented a BCQT scheme using a quantum channel with the nine-QB EN. In this protocol, users can teleport an unknown two-QBS to each other, simultaneously. In the same year, Li et al. [27] introduced a BCQT protocol using a quantum channel with the six-QB cluster state. In this scheme, A can teleport an arbitrary two-QBS to B and B can teleport an arbitrary single-QBS to A under the control of C. Besides, implementation of QT has been presented in some of the works experimentally [28-32] in various quantum systems like the cavity QED system, optical and photonic systems, and ion-trap system.

In this study, a novel BCQT protocol using as a quantum channel with the eight-QB EN is proposed by which the users can teleport an arbitrary EN and a pure two-QBS to each other concurrently under permission of a controller. Also, the probability of guessing the controller's QB by eavesdropper is reduced and the controller (supervisor) can control one of the users or both depending on the intended problem. Also, in the proposed protocol, there is an improve in reducing quantum resources used for preparing the quantum channel into previous works. Then, the proposed protocol is investigated in typical noisy channels including the amplitude-damping noise (ADN) and the phase-damping noise (PDN). Finally, the fidelities of the BCQT process illustrate which they depend on the decoherence noisy rate (DNR) and the amplitude parameter of the initial state (IS). In our scheme, SIQ measurement, controlled-not operation, and appropriate UOs are necessary.

In the following, In Section 2, the proposed BCQT protocol is described. Section 3 presents the preparation and the circuit of the quantum channel. In Section 4, the effect of noise on the proposed protocol is discussed in detail. The comparison between our protocol with previous BCQT works is presented in Section 5. Finally, Sect. 6 concludes the paper.

**Scheme for the presented protocol**

In this protocol, A and B want to transmit an arbitrary EN and a pure two-QBS to each other under the permission of the controller, simultaneously described by Eq. (1).

$$|\Phi\rangle_{A_0 A_1} = \alpha_0|00\rangle + \alpha_1|11\rangle, \qquad (1)$$
$$|\Phi\rangle_{B_0 B_1} = \beta_{00}|00\rangle + \beta_{01}|01\rangle + \beta_{10}|10\rangle + \beta_{11}|11\rangle.$$

Where $|\alpha_0|^2 + |\alpha_1|^2 = 1$ and $|\beta_{00}|^2 + |\beta_{01}|^2 + |\beta_{10}|^2 + |\beta_{11}|^2 = 1$. This protocol include the following steps:

**Step1.** Suppose A, B and C share an eight-QB EN. The structure of the shared channel can change as one of the eight states shown in Table 1 with changing (different distribution) of the dependency of C's QB into $a_0, b_2$, and $b_3$ in the channel. As stated in this Table, C can create and encode the eight different channel by encoding three classical bits, where the QBs $a_0, a_1$, and $a_2$ belong to A, QBs $b_0, b_1, b_2$ and $b_3$ belong to B and Also, QB $c$ belong to C.

Table 1: different channels created using of distribution of C's QB.

| Encoding the various distributions of C's QB | | | The proposed channel $|\Psi\rangle_{a_0 b_0 b_1 a_1 a_2 c b_2 b_3}$ |
|---|---|---|---|
| 0 | 0 | 0 | $\frac{1}{\sqrt{8}}(|00000000\rangle + |00001001\rangle + |00010010\rangle + |00011011\rangle$ $+ |11100000\rangle + |11101001\rangle + |11110010\rangle$ $+ |11111011\rangle),$ |
| 0 | 0 | 1 | $\frac{1}{\sqrt{8}}(|00000000\rangle + |00001101\rangle + |00010010\rangle + |00011111\rangle$ $+ |11100000\rangle + |11101101\rangle + |11110010\rangle$ $+ |11111111\rangle),$ |
| 0 | 1 | 0 | $\frac{1}{\sqrt{8}}(|00000000\rangle + |00001001\rangle + |00010110\rangle + |00011111\rangle$ $+ |11100000\rangle + |11101001\rangle + |11110110\rangle$ $+ |11111111\rangle),$ |
| 0 | 1 | 1 | $\frac{1}{\sqrt{8}}(|00000000\rangle + |00001101\rangle + |00010110\rangle + |00011011\rangle$ $+ |11100000\rangle + |11101101\rangle + |11110110\rangle$ $+ |11111011\rangle),$ |

| 1 | 0 | 0 | $\frac{1}{\sqrt{8}}(|00000000\rangle + |00001001\rangle + |00010010\rangle + |00011011\rangle$ $+ |11100100\rangle + |11101101\rangle + |11110110\rangle$ $+ |11111111\rangle)$, |
|---|---|---|---|
| 1 | 0 | 1 | $\frac{1}{\sqrt{8}}(|00000000\rangle + |00001101\rangle + |00010010\rangle + |00011111\rangle$ $+ |11100100\rangle + |11101001\rangle + |11110110\rangle$ $+ |11111011\rangle)$, |
| 1 | 1 | 0 | $\frac{1}{\sqrt{8}}(|00000000\rangle + |00001001\rangle + |00010110\rangle + |00011111\rangle$ $+ |11100100\rangle + |11101101\rangle + |11110010\rangle$ $+ |11111011\rangle)$, |
| 1 | 1 | 1 | $\frac{1}{\sqrt{8}}(|00000000\rangle + |00001101\rangle + |00010110\rangle + |00011011\rangle$ $+ |11100100\rangle + |11101001\rangle + |11110010\rangle$ $+ |11111111\rangle)$, |

For instance, we assume that the channel shared among A, B and C is the second channel shown in Table 1 with encoding 001 which we described it again in Eq. (2) and we use it for presenting the rest of the proposed protocol.

$$|\Psi\rangle_{a_0 b_0 b_1 a_1 a_2 c b_2 b_3} \quad (2)$$
$$= \frac{1}{\sqrt{8}}(|00000000\rangle + |00001101\rangle + |00010010\rangle$$
$$+ |00011111\rangle$$
$$+ |11100000\rangle + |11101101\rangle + |11110010\rangle +$$
$$|11111111\rangle),$$

The state of the whole system can be expressed by Eq. (3).

$$|\phi\rangle_{a_0 b_0 b_1 a_1 a_2 c b_2 b_3 A_0 A_1 B_0 B_1} = |\Psi\rangle_{a_0 b_0 b_1 a_1 a_2 c b_2 b_3} \otimes |\Phi\rangle_{A_0 A_1} \otimes |\Phi\rangle_{B_0 B_1}. \quad (3)$$

**Step2.** In this step, A and B make a CNOT operation with $A_0, B_0$ and $B_1$ as control QBs and $a_0, b_2$ and $b_3$ as target QBs, respectively. The state will be as Eq. (4).

$|\phi'\rangle_{a_0 b_0 b_1 a_1 a_2 c b_2 b_3 A_0 A_1 B_0 B_1} = \frac{1}{\sqrt{8}} [\alpha_0 \beta_{00}(|00000000\rangle + |00001101\rangle + |00010010\rangle + |00011111\rangle$

$+|11100000\rangle + |11101101\rangle + |11110010\rangle + |11111111\rangle)|0000\rangle$

$+\alpha_0 \beta_{01}(|00000001\rangle + |00001100\rangle + |00010011\rangle + |00011110\rangle$

$+|11100001\rangle + |11101100\rangle + |11110011\rangle + |11111110\rangle)|0001\rangle$

$$
\begin{aligned}
&+\alpha_0\beta_{10}(|00000010\rangle + |00001111\rangle + |00010000\rangle + |00011101\rangle \\
&+|11100010\rangle + |11101111\rangle + |11110000\rangle + |11111101\rangle)|0010\rangle \\
&+\alpha_0\beta_{11}(|00000011\rangle + |00001110\rangle + |00010001\rangle + |00011100\rangle \\
&+|11100011\rangle + |11101110\rangle + |11110001\rangle + |11111100\rangle)|0011\rangle \\
&+\alpha_1\beta_{00}(|10000000\rangle + |10001101\rangle + |10010010\rangle + |10011111\rangle \\
&+|01100000\rangle + |01101101\rangle + |01110010\rangle + |01111111\rangle)|1100\rangle \\
&+\alpha_1\beta_{01}(|10000001\rangle + |10001100\rangle + |10010011\rangle + |10011110\rangle \\
&+|01100001\rangle + |01101100\rangle + |01110011\rangle + |01111110\rangle)|1101\rangle \\
&+\alpha_1\beta_{10}(|10000010\rangle + |10001111\rangle + |10010000\rangle + |10011101\rangle \\
&+|01100010\rangle + |01101111\rangle + |01110000\rangle + |01111101\rangle)|1110\rangle \\
&+\alpha_1\beta_{11}(|10000011\rangle + |10001110\rangle + |10010001\rangle + |10011100\rangle \\
&+|01100011\rangle + |01101110\rangle + |01110001\rangle + |01111100\rangle)|1111\rangle).
\end{aligned}
\tag{4}
$$

**Step3.** A and B do a SIQ measurement in the $Z$-basis on $a_0$, $b_2$ and $b_3$ QBs. According to Table 2, the unmeasured QBs may decrease into one of the eight possible states with the same probability.

Table 2. The Z-basis measurement results of the corresponding collapsed state and users.

| A's results | B's results | The collapsed state of QBs $b_0b_1a_1a_2cA_0A_1B_0B_1$ |
|---|---|---|
| 0 | 00 | $\alpha_0\beta_{00}\|000000000\rangle + \alpha_0\beta_{01}\|000110001\rangle +$ $\alpha_0\beta_{10}\|001000010\rangle + \alpha_0\beta_{11}\|001110011\rangle +$ $\alpha_1\beta_{00}\|110001100\rangle + \alpha_1\beta_{01}\|110111101\rangle +$ $\alpha_1\beta_{10}\|111001110\rangle + \alpha_1\beta_{11}\|111111111\rangle$ |
| 0 | 01 | $\alpha_0\beta_{00}\|000110000\rangle + \alpha_0\beta_{01}\|000000001\rangle +$ $\alpha_0\beta_{10}\|001110010\rangle + \alpha_0\beta_{11}\|001000011\rangle +$ |
| | | $\alpha_1\beta_{00}|110111100\rangle + \alpha_1\beta_{01}|110001101\rangle +$ <br> $\alpha_1\beta_{10}|111111110\rangle + \alpha_1\beta_{11}|111001111\rangle$ |
| 0 | 10 | $\alpha_0\beta_{00}|001000000\rangle + \alpha_0\beta_{01}|001110001\rangle +$ <br> $\alpha_0\beta_{10}|000000010\rangle + \alpha_0\beta_{11}|000110011\rangle +$ <br> $\alpha_1\beta_{00}|111001100\rangle + \alpha_1\beta_{01}|111111101\rangle +$ <br> $\alpha_1\beta_{10}|110001110\rangle + \alpha_1\beta_{11}|110111111\rangle$ |
| 0 | 11 | $\alpha_0\beta_{00}|001110000\rangle + \alpha_0\beta_{01}|001000001\rangle +$ <br> $\alpha_0\beta_{10}|000110010\rangle + \alpha_0\beta_{11}|000000011\rangle +$ <br> $\alpha_1\beta_{00}|111111100\rangle + \alpha_1\beta_{01}|111001101\rangle +$ <br> $\alpha_1\beta_{10}|110111110\rangle + \alpha_1\beta_{11}|110001111\rangle$ |
| 1 | 00 | $\alpha_0\beta_{00}|110000000\rangle + \alpha_0\beta_{01}|110110001\rangle +$ <br> $\alpha_0\beta_{10}|111000010\rangle + \alpha_0\beta_{11}|111110011\rangle +$ <br> $\alpha_1\beta_{00}|000001100\rangle + \alpha_1\beta_{01}|000111101\rangle +$ <br> $\alpha_1\beta_{10}|001001110\rangle + \alpha_1\beta_{11}|001111111\rangle$ |
| 1 | 01 | $\alpha_0\beta_{00}|110110000\rangle + \alpha_0\beta_{01}|110000001\rangle +$ <br> $\alpha_0\beta_{10}|111110010\rangle + \alpha_0\beta_{11}|111000011\rangle +$ <br> $\alpha_1\beta_{00}|000111100\rangle + \alpha_1\beta_{01}|000001101\rangle +$ <br> $\alpha_1\beta_{10}|001111110\rangle + \alpha_1\beta_{11}|001001111\rangle$ |
| 1 | 10 | $\alpha_0\beta_{00}|111000000\rangle + \alpha_0\beta_{01}|111110001\rangle +$ <br> $\alpha_0\beta_{10}|110000010\rangle + \alpha_0\beta_{11}|110110011\rangle +$ <br> $\alpha_1\beta_{00}|001001100\rangle + \alpha_1\beta_{01}|001111101\rangle +$ <br> $\alpha_1\beta_{10}|000001110\rangle + \alpha_1\beta_{11}|000111111\rangle$ |
| 1 | 11 | $\alpha_0\beta_{00}|111110000\rangle + \alpha_0\beta_{01}|111000001\rangle +$ <br> $\alpha_0\beta_{10}|110110010\rangle + \alpha_0\beta_{11}|110000011\rangle +$ <br> $\alpha_1\beta_{00}|001111100\rangle + \alpha_1\beta_{01}|001001101\rangle +$ <br> $\alpha_1\beta_{10}|000111110\rangle + \alpha_1\beta_{11}|000001111\rangle$ |

**Step4.** In this step A and B notify the Z-basis measurement results to each other. Then they apply $X$ UO on QBs $b_0, b_1, a_1$ and $a_2$ as shown in Table 3. The state of the unmeasured QBs will be converted to the same form:

$$\alpha_0\beta_{00}|000000000\rangle + \alpha_0\beta_{01}|000110001\rangle + \alpha_0\beta_{10}|001000010\rangle + \quad (5)$$
$$\alpha_0\beta_{11}|001110011\rangle + \alpha_1\beta_{00}|110001100\rangle + \alpha_1\beta_{01}|110111101\rangle +$$
$$\alpha_1\beta_{10}|111001110\rangle + \alpha_1\beta_{11}|111111111\rangle$$

Table 3. Applying $X$ UO

| A's Result | B's Result | UO on $(b0)(b1)(a1)(a2)$ |
|---|---|---|
| 0 | 00 | $I \otimes I \otimes I \otimes I$ |
| 0 | 01 | $I \otimes I \otimes I \otimes X$ |
| 0 | 10 | $I \otimes I \otimes X \otimes I$ |
| 0 | 11 | $I \otimes I \otimes X \otimes X$ |
| 1 | 00 | $X \otimes X \otimes I \otimes I$ |
| 1 | 01 | $X \otimes X \otimes I \otimes X$ |
| 1 | 10 | $X \otimes X \otimes X \otimes I$ |
| 1 | 11 | $X \otimes X \otimes X \otimes X$ |

**Step5.** SIQ measurements are applied in the $X$-basis on sending QBs ($A_0, A_1, B_0$ and $B_1$) by A and B. As shown in Table 4, the other QBs collapsed to one of 16 possible states with same probability.

Table 4. The measurement results based on $X$ and collapsed states.

| A's Result | B's Result | The collapsed state of QBs $(b_0)(b_1)(a_1)(a_2)(c)$ |
|---|---|---|
| + + | + + | $\alpha_0\beta_{00}\|00000\rangle + \alpha_0\beta_{01}\|00011\rangle + \alpha_0\beta_{10}\|00100\rangle$ <br> $+ \alpha_0\beta_{11}\|00111\rangle + \alpha_1\beta_{00}\|11000\rangle$ <br> $+ \alpha_1\beta_{01}\|11011\rangle + \alpha_1\beta_{10}\|11100\rangle$ <br> $+ \alpha_1\beta_{11}\|11111\rangle$ |
| + + | + − | $\alpha_0\beta_{00}\|00000\rangle - \alpha_0\beta_{01}\|00011\rangle + \alpha_0\beta_{10}\|00100\rangle$ <br> $- \alpha_0\beta_{11}\|00111\rangle + \alpha_1\beta_{00}\|11000\rangle$ <br> $- \alpha_1\beta_{01}\|11011\rangle + \alpha_1\beta_{10}\|11100\rangle$ <br> $- \alpha_1\beta_{11}\|11111\rangle$ |
| + + | − + | $\alpha_0\beta_{00}\|00000\rangle + \alpha_0\beta_{01}\|00011\rangle - \alpha_0\beta_{10}\|00100\rangle$ <br> $- \alpha_0\beta_{11}\|00111\rangle + \alpha_1\beta_{00}\|11000\rangle$ <br> $+ \alpha_1\beta_{01}\|11011\rangle - \alpha_1\beta_{10}\|11100\rangle$ <br> $- \alpha_1\beta_{11}\|11111\rangle$ |
| + + | − − | $\alpha_0\beta_{00}\|00000\rangle - \alpha_0\beta_{01}\|00011\rangle - \alpha_0\beta_{10}\|00100\rangle$ <br> $+ \alpha_0\beta_{11}\|00111\rangle + \alpha_1\beta_{00}\|11000\rangle$ <br> $- \alpha_1\beta_{01}\|11011\rangle - \alpha_1\beta_{10}\|11100\rangle$ <br> $+ \alpha_1\beta_{11}\|11111\rangle$ |
| $+\,-$ | $+\,+$ | $\alpha_0\beta_{00}\|00000\rangle + \alpha_0\beta_{01}\|00011\rangle + \alpha_0\beta_{10}\|00100\rangle + \alpha_0\beta_{11}\|00111\rangle + -\alpha_1\beta_{00}\|11000\rangle - \alpha_1\beta_{01}\|11011\rangle - \alpha_1\beta_{10}\|11100\rangle - \alpha_1\beta_{11}\|11111\rangle$ |
| $+\,-$ | $+\,-$ | $\alpha_0\beta_{00}\|00000\rangle - \alpha_0\beta_{01}\|00011\rangle + \alpha_0\beta_{10}\|00100\rangle - \alpha_0\beta_{11}\|00111\rangle + -\alpha_1\beta_{00}\|11000\rangle + \alpha_1\beta_{01}\|11011\rangle - \alpha_1\beta_{10}\|11100\rangle + \alpha_1\beta_{11}\|11111\rangle$ |
| $+\,-$ | $-\,+$ | $\alpha_0\beta_{00}\|00000\rangle + \alpha_0\beta_{01}\|00011\rangle - \alpha_0\beta_{10}\|00100\rangle - \alpha_0\beta_{11}\|00111\rangle - \alpha_1\beta_{00}\|11000\rangle - \alpha_1\beta_{01}\|11011\rangle + \alpha_1\beta_{10}\|11100\rangle + \alpha_1\beta_{11}\|11111\rangle$ |
| $+\,-$ | $-\,-$ | $\alpha_0\beta_{00}\|00000\rangle - \alpha_0\beta_{01}\|00011\rangle - \alpha_0\beta_{10}\|00100\rangle + \alpha_0\beta_{11}\|00111\rangle - \alpha_1\beta_{00}\|11000\rangle + \alpha_1\beta_{01}\|11011\rangle + \alpha_1\beta_{10}\|11100\rangle - \alpha_1\beta_{11}\|11111\rangle$ |
| $-\,+$ | $+\,+$ | $\alpha_0\beta_{00}\|00000\rangle + \alpha_0\beta_{01}\|00011\rangle + \alpha_0\beta_{10}\|00100\rangle + \alpha_0\beta_{11}\|00111\rangle - \alpha_1\beta_{00}\|11000\rangle - \alpha_1\beta_{01}\|11011\rangle - \alpha_1\beta_{10}\|11100\rangle - \alpha_1\beta_{11}\|11111\rangle$ |
| $-\,+$ | $+\,-$ | $\alpha_0\beta_{00}\|00000\rangle - \alpha_0\beta_{01}\|00011\rangle + \alpha_0\beta_{10}\|00100\rangle - \alpha_0\beta_{11}\|00111\rangle - \alpha_1\beta_{00}\|11000\rangle + \alpha_1\beta_{01}\|11011\rangle - \alpha_1\beta_{10}\|11100\rangle + \alpha_1\beta_{11}\|11111\rangle$ |
| $-\,+$ | $-\,+$ | $\alpha_0\beta_{00}\|00000\rangle + \alpha_0\beta_{01}\|00011\rangle - \alpha_0\beta_{10}\|00100\rangle - \alpha_0\beta_{11}\|00111\rangle - \alpha_1\beta_{00}\|11000\rangle - \alpha_1\beta_{01}\|11011\rangle + \alpha_1\beta_{10}\|11100\rangle + \alpha_1\beta_{11}\|11111\rangle$ |
| $-\,+$ | $-\,-$ | $\alpha_0\beta_{00}\|00000\rangle - \alpha_0\beta_{01}\|00011\rangle - \alpha_0\beta_{10}\|00100\rangle + \alpha_0\beta_{11}\|00111\rangle - \alpha_1\beta_{00}\|11000\rangle + \alpha_1\beta_{01}\|11011\rangle + \alpha_1\beta_{10}\|11100\rangle - \alpha_1\beta_{11}\|11111\rangle$ |
| $-\,-$ | $+\,+$ | $\alpha_0\beta_{00}\|00000\rangle + \alpha_0\beta_{01}\|00011\rangle + \alpha_0\beta_{10}\|00100\rangle + \alpha_0\beta_{11}\|00111\rangle + \alpha_1\beta_{00}\|11000\rangle + \alpha_1\beta_{01}\|11011\rangle + \alpha_1\beta_{10}\|11100\rangle + \alpha_1\beta_{11}\|11111\rangle$ |
| $--$ | $+-$ | $\alpha_0\beta_{00}\|00000\rangle - \alpha_0\beta_{01}\|00011\rangle + \alpha_0\beta_{10}\|00100\rangle$ $- \alpha_0\beta_{11}\|00111\rangle + \alpha_1\beta_{00}\|11000\rangle$ $- \alpha_1\beta_{01}\|11011\rangle + \alpha_1\beta_{10}\|11100\rangle$ $- \alpha_1\beta_{11}\|11111\rangle$ |
| $--$ | $-+$ | $\alpha_0\beta_{00}\|00000\rangle + \alpha_0\beta_{01}\|00011\rangle - \alpha_0\beta_{10}\|00100\rangle$ $- \alpha_0\beta_{11}\|00111\rangle + \alpha_1\beta_{00}\|11000\rangle$ $+ \alpha_1\beta_{01}\|11011\rangle - \alpha_1\beta_{10}\|11100\rangle$ $- \alpha_1\beta_{11}\|11111\rangle$ |
| $--$ | $--$ | $\alpha_0\beta_{00}\|00000\rangle - \alpha_0\beta_{01}\|00011\rangle - \alpha_0\beta_{10}\|00100\rangle$ $+ \alpha_0\beta_{11}\|00111\rangle + \alpha_1\beta_{00}\|11000\rangle$ $- \alpha_1\beta_{01}\|11011\rangle - \alpha_1\beta_{10}\|11100\rangle$ $+ \alpha_1\beta_{11}\|11111\rangle$ |

**Step6.** A and B notify their measurement results to each other. Then, they apply $Z$ UO to their unmeasured QBs ($b_0, b_1, a_1$ and $a_2$) as shown in Table 5. The state of the unmeasured QBs will be converted to the same form (6).

$$\alpha_0\beta_{00}|00000\rangle + \alpha_0\beta_{01}|00011\rangle + \alpha_0\beta_{10}|00100\rangle + \alpha_0\beta_{11}|00111\rangle \qquad (6)$$
$$+ \alpha_1\beta_{00}|11000\rangle + \alpha_1\beta_{01}|11011\rangle + \alpha_1\beta_{10}|11100\rangle$$
$$+ \alpha_1\beta_{11}|11111\rangle$$

Table 5. Applying $Z$ UO.

| A's Result | B's Result | UO on $(b_0)(b_1)(a_1)(a_2)$ |
|---|---|---|
| $++$ | $++$ | $I\otimes I\otimes I\otimes I$ |
| $++$ | $+-$ | $I\otimes I\otimes I\otimes Z$ |
| $++$ | $-+$ | $I\otimes I\otimes Z\otimes I$ |
| $++$ | $--$ | $I\otimes I\otimes Z\otimes Z$ |
| $+-$ | $++$ | $I\otimes Z\otimes I\otimes I$ |
| $+-$ | $+-$ | $I\otimes Z\otimes I\otimes Z$ |
| $+-$ | $-+$ | $I\otimes Z\otimes Z\otimes I$ |
| $+-$ | $--$ | $I\otimes Z\otimes Z\otimes Z$ |
| $-+$ | $++$ | $Z\otimes I\otimes I\otimes I$ |
| $-+$ | $+-$ | $Z\otimes I\otimes I\otimes Z$ |
| $-+$ | $-+$ | $Z\otimes I\otimes Z\otimes I$ |
| $-+$ | $--$ | $Z\otimes I\otimes Z\otimes Z$ |
| − − | + + | $Z\otimes Z\otimes I\otimes I$ |
| − − | + − | $Z\otimes Z\otimes I\otimes Z$ |
| − − | − + | $Z\otimes Z\otimes Z\otimes I$ |
| − − | − − | $Z\otimes Z\otimes Z\otimes Z$ |

**Step7.** C notifies distribution status of his QB to A and B with three classical bits as shown in Table 1. Then, he measures her QB in *X*-basis and tells to A and B his result. If C's measured result is $|+\rangle$ ($|-\rangle$), then, the state of other QBs is as (7) or (8), respectively. The measurement results of C's QB with corresponding unitary operations (UOs) applied by A and B are shown in Table 6 for the different channels in Table 1.

$$\alpha_0\beta_{00}|0000\rangle + \alpha_0\beta_{01}|0001\rangle + \alpha_0\beta_{10}|0010\rangle + \alpha_0\beta_{11}|0011\rangle + \alpha_1\beta_{00}|1100\rangle \quad (7)$$
$$+ \alpha_1\beta_{01}|1101\rangle + \alpha_1\beta_{10}|1110\rangle + \alpha_1\beta_{11}|1111\rangle$$

$$\alpha_0\beta_{00}|0000\rangle - \alpha_0\beta_{01}|0001\rangle + \alpha_0\beta_{10}|0010\rangle - \alpha_0\beta_{11}|0011\rangle + \alpha_1\beta_{00}|1100\rangle \quad (8)$$
$$- \alpha_1\beta_{01}|1101\rangle + \alpha_1\beta_{10}|1110\rangle - \alpha_1\beta_{11}|1111\rangle$$

Table 6. Applying *Z* UO for the deferent channel showed in Table 1.

| Coding bits to show the different channels | C's Results | The collapsed state of QBs $b_0, b_1, a_1, a_2$ | UO on $b_0, b_1, a_1, a_2$ |
|---|---|---|---|
| 000 | $|+\rangle$ | $|0000\rangle + |0001\rangle + |0010\rangle + |0011\rangle + |1100\rangle + |1101\rangle + |1110\rangle + |1111\rangle$ | $I\otimes I\otimes I\otimes I$ |
| 000 | $|-\rangle$ | $|0000\rangle + |0001\rangle + |0010\rangle + |0011\rangle + |1100\rangle + |1101\rangle + |1110\rangle + |1111\rangle$ | $I\otimes I\otimes I\otimes I$ |
| 001 | $|+\rangle$ | $|0000\rangle + |0001\rangle + |0010\rangle + |0011\rangle + |1100\rangle + |1101\rangle + |1110\rangle + |1111\rangle$ | $I\otimes I\otimes I\otimes I$ |
| 001 | $|-\rangle$ | $|0000\rangle - |0001\rangle + |0010\rangle - |0011\rangle + |1100\rangle - |1101\rangle + |1110\rangle - |1111\rangle$ | $I\otimes I\otimes I\otimes Z$ |

| | | | |
|---|---|---|---|
| 010 | $\|+\rangle$ | $\|0000\rangle + \|0001\rangle + \|0010\rangle + \|0011\rangle + \|1100\rangle + \|1101\rangle + \|1110\rangle + \|1111\rangle$ | $I \otimes I \otimes I \otimes I$ |
| | $\|-\rangle$ | $\|0000\rangle + \|0001\rangle - \|0010\rangle - \|0011\rangle + \|1100\rangle + \|1101\rangle - \|1110\rangle - \|1111\rangle$ | $I \otimes I \otimes Z \otimes I$ |
| 011 | $\|+\rangle$ | $\|0000\rangle + \|0001\rangle + \|0010\rangle + \|0011\rangle + \|1100\rangle + \|1101\rangle + \|1110\rangle + \|1111\rangle$ | $I \otimes I \otimes I \otimes I$ |
| | $\|-\rangle$ | $\|0000\rangle - \|0001\rangle - \|0010\rangle + \|0011\rangle + \|1100\rangle - \|1101\rangle - \|1110\rangle + \|1111\rangle$ | $I \otimes I \otimes Z \otimes Z$ |
| 100 | $\|+\rangle$ | $\|0000\rangle + \|0001\rangle + \|0010\rangle + \|0011\rangle + \|1100\rangle + \|1101\rangle + \|1110\rangle + \|1111\rangle$ | $I \otimes I \otimes I \otimes I$ |
| | $\|-\rangle$ | $\|0000\rangle + \|0001\rangle + \|0010\rangle + \|0011\rangle - \|1100\rangle - \|1101\rangle - \|1110\rangle - \|1111\rangle$ | $I \otimes Z \otimes I \otimes I$ OR $Z \otimes I \otimes I \otimes I$ |
| 101 | $\|+\rangle$ | $\|0000\rangle + \|0001\rangle + \|0010\rangle + \|0011\rangle + \|1100\rangle + \|1101\rangle + \|1110\rangle + \|1111\rangle$ | $I \otimes I \otimes I \otimes I$ |
| | $\|-\rangle$ | $\|0000\rangle - \|0001\rangle + \|0010\rangle - \|0011\rangle - \|1100\rangle + \|1101\rangle - \|1110\rangle + \|1111\rangle$ | $I \otimes Z \otimes I \otimes Z$ OR $Z \otimes I \otimes I \otimes Z$ |
| 110 | $\|+\rangle$ | $\|0000\rangle + \|0001\rangle + \|0010\rangle + \|0011\rangle + \|1100\rangle + \|1101\rangle + \|1110\rangle + \|1111\rangle$ | $I \otimes I \otimes I \otimes I$ |
| | $\|-\rangle$ | $\|0000\rangle + \|0001\rangle - \|0010\rangle - \|0011\rangle - \|1100\rangle - \|1101\rangle + \|1110\rangle + \|1111\rangle$ | $I \otimes Z \otimes Z \otimes I$ OR $Z \otimes I \otimes Z \otimes I$ |
| 111 | $\|+\rangle$ | $\|0000\rangle + \|0001\rangle + \|0010\rangle + \|0011\rangle + \|1100\rangle + \|1101\rangle + \|1110\rangle + \|1111\rangle$ | $I \otimes I \otimes I \otimes I$ |

| | $\lvert - \rangle$ | $\lvert 0000\rangle - \lvert 0001\rangle - \lvert 0010\rangle + \lvert 0011\rangle - \lvert 1100\rangle + \lvert 1101\rangle + \lvert 1110\rangle - \lvert 1111\rangle$ | $I \otimes Z \otimes Z \otimes Z$ OR $Z \otimes I \otimes Z \otimes Z$ |
|---|---|---|---|

**Step8.** According to C's results, A and B apply $Z$ UO as shown in Table 6. In final, A and B can reconstruct the transmitted states again as (9) and (10). And the BCQT is successfully finished.

$$\lvert \Phi \rangle_{A_0 A_1} = \beta_{00}\lvert 00\rangle + \beta_{01}\lvert 01\rangle + \beta_{10}\lvert 10\rangle + \beta_{11}\lvert 11\rangle \tag{9}$$

$$\lvert \Phi \rangle_{B_0 B_1} = \alpha_0 \lvert 00\rangle + \alpha_1 \lvert 11\rangle \tag{10}$$

**Preparation of eight-qubit entangled state (QB ENS)**

The proposed quantum channel (eight-QB EN) is practically feasible as shown in Fig. 1. As shown in this figure, the quantum circuit of the proposed channel can be created by utilizing three Hadamard gates and using four to seven CNOT gates.

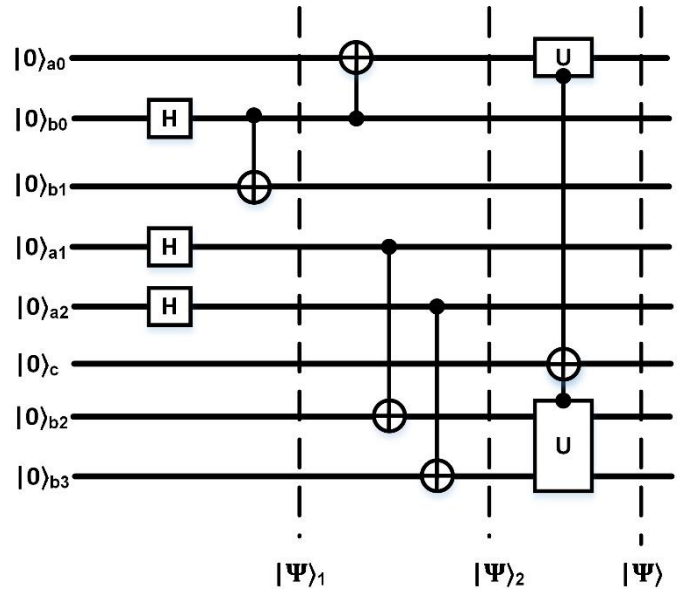

Fig. 1: Quantum circuit for preparing eight-QB quantum channel.

The steps of creating a channel are explained in details as the following:
The first, IS is prepared with zero states as Eq.11.

$$\lvert \Psi \rangle_0 = \lvert 0 \rangle_{a0} \otimes \lvert 0 \rangle_{b0} \otimes \lvert 0 \rangle_{b1} \otimes \lvert 0 \rangle_{a1} \otimes \lvert 0 \rangle_{a2} \otimes \lvert 0 \rangle_c \otimes \lvert 0 \rangle_{b2} \otimes \lvert 0 \rangle_{b3}. \tag{11}$$

After applying Hadamard gates, one CNOT gate is applied with QB b0 as control QB and b1 as target QB. Then, the whole state of system is as the following:

$$|\Psi\rangle_1 = \quad (12)$$

$$CNOT(b0, b1) \; (|0\rangle_{a0} \otimes \frac{(|0\rangle+|1\rangle)}{\sqrt{2}}_{b0} \otimes |0\rangle_{b1} \otimes \frac{(|0\rangle+|1\rangle)}{\sqrt{2}}_{a1} \otimes \frac{(|0\rangle+|1\rangle)}{\sqrt{2}}_{a2} \otimes |0\rangle_c \otimes |0\rangle_{b2} \otimes |0\rangle_{b3})$$

$$= \frac{1}{\sqrt{8}}(|00000000\rangle + |00001000\rangle + |00010000\rangle + |00011000\rangle + |01100000\rangle + |01101000\rangle + |01110000\rangle + |01111000\rangle)_{a_0 b_0 b_1 a_1 a_2 c b_2 b_3}.$$

In the next step, three CNOT gates are applied as that QBs $b0$, $a1$ and $a2$ are Control QBs and QBs $a0, b2$ and $b3$ target QBs, respectively. Then, the state of all the eight QBs becomes as Eq. (13).

$$|\Psi\rangle_1 = \frac{1}{\sqrt{8}}(|00000000\rangle + |00001001\rangle + |00010010\rangle + |00011011\rangle + |11100000\rangle + |11101001\rangle + |11110010\rangle + |11111011\rangle)_{a_0 b_0 b_1 a_1 a_2 c b_2 b_3} \quad (13)$$

Finally, the controller can apply CNOT gates by $U$ function so that he can consider each combination of QBs $a0, b2$ and $b3$ as control and QB $c$ as the target. So, he can create different eight states as shown in Table 1 and type of used combination can be encoded by three classical bits according to sequence $a0, b2, b3$. For example, if he will create a state in Eq. (2), then he needs to apply one CNOT gate with QB $b3$ as control and QB $c$ as the target. So, the proposed channel can be created as Eq. (14).

$$|\Psi\rangle = \frac{1}{\sqrt{8}}(|00000000\rangle + |00001101\rangle + |00010010\rangle + |00011111\rangle + |11100000\rangle + |11101101\rangle + |11110010\rangle + |11111111\rangle)_{a_0 b_0 b_1 a_1 a_2 c b_2 b_3} \quad (14)$$

As stated above, an eight-QBS can be prepared and used as a quantum channel. Until now, the SIQ UOs and single QB measurements have already been used in various quantum systems [28-29-30]. So, the proposed scheme will be implemented in quantum information technology with advances in the future.

**Effects of channel noises on the proposed protocol**

In this section, the effect of two models of environment noise includes of an AD and PD noisy environment on proposed BCQT process are discussed. These two environment noise can be determined by Kraus operators [33] shown in EQBS. (15) and (16), respectively.

$$E_0^A = \begin{bmatrix} 1 & 0 \\ 0 & \sqrt{1-\eta_A} \end{bmatrix}, E_1^A = \begin{bmatrix} 0 & \sqrt{\eta_A} \\ 0 & 0 \end{bmatrix}. \quad (15)$$

$$E_0^P = \sqrt{1-\eta_P}I, E_1^P = \sqrt{\eta_P}\begin{bmatrix} 1 & 0 \\ 0 & 0 \end{bmatrix}, E_2^P = \sqrt{\eta_P}\begin{bmatrix} 0 & 0 \\ 0 & 1 \end{bmatrix}. \quad (16)$$

Where $\eta_A$ ($0 \leq \eta_A \leq 1$) and $\eta_P$ ($0 \leq \eta_P \leq 1$) are the DRs (the error occurring probability of a quantum state (QS) in the corresponding channel when a travel QB passes through it) for the a ADN and the PDN, respectively. Also, $I$ is the identity matrix in the Hilbert space of $C_{2\times 2}$. The channel proposed in the previous section ($|\Psi\rangle$) is a pure state. The corresponding density matrix can be described as $\rho = |\Psi\rangle\langle\Psi|$. So, the effect of the noise described by (15) or (16) on the density operator $\rho$ can be stated as Eq. (17).

$$\xi^r(\rho) = \sum_m (E_m^{ra0})(E_m^{ra1})(E_m^{ra2})(E_m^{rb0})(E_m^{rb1})(E_m^{rb2})(E_m^{rb3})\rho \quad (17)$$
$$(E_m^{ra0})^\dagger(E_m^{ra1})^\dagger(E_m^{ra2})^\dagger(E_m^{rb0})^\dagger(E_m^{rb1})^\dagger(E_m^{rb2})^\dagger(E_m^{rb3})^\dagger$$

Where $r \in \{A, P\}$. For $r = A$, i.e, for ADN $m \in \{0,1\}$, while for $r = P$, i.e., for PDN $m \in \{0,1,2\}$, and the superscripts $ai$ and $bi$ represent the operator $E$ act on which QB. $\xi$ denotes a quantum operation which maps from $\rho$ to $\xi^r(\rho)$ due to the noise. In Eq. (17), we suppose that QBs belong to A and B are affected by noisy environments (NEs) due to that controller (C) create the channel and these QBs are transmitted through the NE by the controller (C) to A and B. But, QB $c$ is considered without effect of the NE due to belonging to controller (C) and is not transmitted in the channel. Also, it is considered that both the QBs sent to A and B are affected by the same Kraus operator. In these two-noise environments, the quantum channel would become a mixed state as shown in Eq. (18) and (19).

$$\xi^A(\rho) = \frac{1}{8}\Big\{\big[|00000000\rangle + (1-\eta_A)|00001101\rangle + (1-\eta_A)|00010010\rangle + (1-\eta_A)^2|00011111\rangle + \sqrt{(1-\eta_A)^3}|11100000\rangle + \sqrt{(1-\eta_A)^5}|11101101\rangle + \sqrt{(1-\eta_A)^5}|11110010\rangle + \sqrt{(1-\eta_A)^7}|11111111\rangle\big] \times \big[\langle 00000000| + (1-\eta_A)\langle 00001101| + (1-\eta_A)\langle 00010010| + (1-\eta_A)^2\langle 00011111| + \sqrt{(1-\eta_A)^3}\langle 11100000| + \sqrt{(1-\eta_A)^5}\langle 11101101| + \sqrt{(1-\eta_A)^5}\langle 11110010| + \sqrt{(1-\eta_A)^7}\langle 11111111|\big] + \eta_A^7|00000100\rangle\langle 00000100|\Big\}.$$

(18)

And

$$\xi^P(\rho) = \frac{1}{8}\{(1-\eta_P)^7[|00000000\rangle + |00001101\rangle + |00010010\rangle + |00011111\rangle +$$

$$|11100000\rangle + |11101101\rangle + |11110010\rangle + |11111111\rangle] \times [\langle 00000000| +$$

$$\langle 00001101| + \langle 00010010| + \langle 00011111| + \langle 11100000| + \langle 11101101| +$$

$$\langle 11110010| + \langle 11111111|] + \eta_P^7|00000000\rangle\langle 00000000| +$$

$$\eta_P^7|11111111\rangle\langle 11111111|\}.$$

(19)

According to A's and B's and C's measured results, A and B can make apt operation on own QBs to recover the original state. Then, the final state can be represented as a density matrix as shown in Eq. (20)

$$\rho_{out}^r = Tr_{A_0A_1B_0B_1a_0cb_2b_3}\{U[\rho_{A_0A_1}\otimes\rho_{B_0B_1}\otimes\xi^r(\rho)]U^\dagger\}, \tag{20}$$

Where $Tr_{A_0A_1B_0B_1a_0cb_2b_3}$ is the partial trace over QBs $(A_0, A_1, B_0, B_1, a_0, c, b_2, b_3)$ and $U$ is a UO to explain the BQCT process, which is written by

$$U = \{I_{A_0A_1a_0} \otimes I_{B_0B_1b_2b_3} \otimes I_c \otimes \sigma_{b_0b_1a_1a_2}^{lmnopqst}\}$$

$$\{I_{A_0A_1a_0} \otimes I_{B_0B_1b_2b_3} \otimes |\phi\rangle_c^t\langle\phi|_c^t \otimes I_{b_0b_1a_1a_2}\}$$

$$\{|\phi\rangle_{A_0}^s\langle\phi|_{A_0}^s \otimes |\phi\rangle_{A_1}^q\langle\phi|_{A_1}^q \otimes |\phi\rangle_{B_0}^p\langle\phi|_{B_0}^p \otimes |\phi\rangle_{B_1}^o\langle\phi|_{B_1}^o \otimes I_{a_0} \otimes I_{b_2b_3} \otimes I_c \otimes I_{b_0b_1a_1a_2}\}$$

$$\{I_{A_0A_1} \otimes I_{B_0B_1} \otimes |\phi\rangle_{a_0}^n\langle\phi|_{a_0}^n \otimes |\phi\rangle_{b_2}^m\langle\phi|_{b_2}^m \otimes |\phi\rangle_{b_3}^l\langle\phi|_{b_3}^l \otimes I_c \otimes I_{b_0b_1a_1a_2}\}$$

$$\{U_{A_0a_0} \otimes U_{B_0b_2} \otimes U_{B_1b_3} \otimes I_{A_1} \otimes I_c \otimes I_{b_0b_1a_1a_2}\}.$$

(21)

Where $l, m, n, o, p, q, s, t \in \{1,2\}$, with $U_{i,j}$ stands for controlled_not operations on the QBs $A_0, B_0$ and $B_1$ as control and QBs $a_0, b_2$ and $b_3$ as target, $|\phi\rangle_{a_0}^n\langle\phi|_{a_0}^n$, $|\phi\rangle_{A_0}^s\langle\phi|_{A_0}^s$ and $|\phi\rangle_{A_1}^q\langle\phi|_{A_1}^q$ denote A's single-QBS measurement results, $|\phi\rangle_{b_2}^m\langle\phi|_{b_2}^m$, $|\phi\rangle_{b_3}^l\langle\phi|_{b_3}^l$, $|\phi\rangle_{B_0}^p\langle\phi|_{B_0}^p$ and $|\phi\rangle_{B_1}^o\langle\phi|_{B_1}^o$ denote B's single-QBS measurement results, and $|\phi\rangle_c^t\langle\phi|_c^t$ shows C's single-QBS measurement result. $\sigma_{b_0b_1a_1a_2}^{lmnopqst}$ is A's and B's recover operation depending on A's and B's and C's measurement results.

Given that the choice, we may have to take the final QS $\rho_{out}^r$ which is the product of the QSs generated on the side of the receivers A and B in a NE. On the other hand, where A would have QB $(\beta_{00}|00\rangle + \beta_{01}|01\rangle + \beta_{10}|10\rangle + \beta_{11}|11\rangle)_{a_1a_2}$ in her possession and B would have

$(\alpha_0|00\rangle + \alpha_1|11\rangle)_{b_0 b_1}$ in his possession, the expected final state in the absence of noise is a product state. As we know, in the ideal situation would be $|\Phi\rangle = (\beta_{00}|00\rangle + \beta_{01}|01\rangle + \beta_{10}|10\rangle + \beta_{11}|11\rangle)_{a_1 a_2} \otimes (\alpha_0|00\rangle + \alpha_1|11\rangle)_{b_0 b_1}$.

By comparing the QS $\rho^r_{out}$ in the NE with the state $|\Phi\rangle$ by using fidelity the effect of noise can be shown as follow:

$$F = \langle\Phi|\rho^r_{out}|\Phi\rangle. \tag{22}$$

According to (20) and (21), get the resultant state $\rho^r_{out}$ from (18) and (19) is easy and the $\rho^r_{out}$ is related to three participants' measurement results $l, m, n, o, p, q, s,$ and $t$. However, after calculation, we get that output state $\rho^r_{out}$ is independent of three participants' measurement results. And the $\rho^r_{out}$ is shown as follows.

$$\rho^A_{out} = \{[\alpha_0\beta_{00}|0000\rangle + (1-\eta_A)\alpha_0\beta_{01}|0001\rangle + (1-\eta_A)\alpha_0\beta_{10}|0010\rangle + (1-\eta_A)^2\alpha_0\beta_{11}|0011\rangle + \sqrt{(1-\eta_A)^3}\alpha_1\beta_{00}|1100\rangle + \sqrt{(1-\eta_A)^5}\alpha_1\beta_{01}|1101\rangle + \sqrt{(1-\eta_A)^5}\alpha_1\beta_{10}|1110\rangle + \sqrt{(1-\eta_A)^7}\alpha_1\beta_{11}|1111\rangle] \times [\alpha_0\beta_{00}\langle 0000| + (1-\eta_A)\alpha_0\beta_{01}\langle 0001| + (1-\eta_A)\alpha_0\beta_{10}\langle 0010| + (1-\eta_A)^2\alpha_0\beta_{11}\langle 0011| + \sqrt{(1-\eta_A)^3}\alpha_1\beta_{00}\langle 1100| + \sqrt{(1-\eta_A)^5}\alpha_1\beta_{01}\langle 1101| + \sqrt{(1-\eta_A)^5}\alpha_1\beta_{10}\langle 1110| + \sqrt{(1-\eta_A)^7}\alpha_1\beta_{11}\langle 1111|] + \eta_A^7\alpha_1^2\beta_{11}^2|0000\rangle\langle 0000|\}.$$

(23)

And

$$\rho^P_{out} = \{(1-\eta_P)^7[\alpha_0\beta_{00}|0000\rangle + \alpha_0\beta_{01}|0001\rangle + \alpha_0\beta_{10}|0010\rangle + \alpha_0\beta_{11}|0011\rangle + \alpha_1\beta_{00}|1100\rangle + \alpha_1\beta_{01}|1101\rangle + \alpha_1\beta_{10}|1110\rangle + \alpha_1\beta_{11}|1111\rangle] \times [\alpha_0\beta_{00}\langle 0000| + \alpha_0\beta_{01}\langle 0001| + \alpha_0\beta_{10}\langle 0010| + \alpha_0\beta_{11}\langle 0011| + \alpha_1\beta_{00}\langle 1100| + \alpha_1\beta_{01}\langle 1101| + \alpha_1\beta_{10}\langle 1110| + \alpha_1\beta_{11}\langle 1111|] + \eta_P^7\alpha_0^2\beta_{00}^2|0000\rangle\langle 0000| + \eta_P^7\alpha_1^2\beta_{11}^2|1111\rangle\langle 1111|\}.$$

(24)

Using (22) and (23), the fidelity of the QS teleportation obtained by utilizing the proposed BQCT under ADN as below:

$$F^A = \{[\alpha_0^2\beta_{00}^2 + (1-\eta_A)\alpha_0^2\beta_{01}^2 + (1-\eta_A)\alpha_0^2\beta_{10}^2 + (1-\eta_A)^2\alpha_0^2\beta_{11}^2 + \sqrt{(1-\eta_A)^3}\alpha_1^2\beta_{00}^2 + \sqrt{(1-\eta_A)^5}\alpha_1^2\beta_{01}^2 + \sqrt{(1-\eta_A)^5}\alpha_1^2\beta_{10}^2 + \sqrt{(1-\eta_A)^7}\alpha_1^2\beta_{11}^2]^2 + \eta_A^7\alpha_0^2\alpha_1^2\beta_{00}^2\beta_{11}^2\}. \tag{25}$$

Similarly, by using (22) and (24), the fidelity of the QS teleportation obtained by utilizing the proposed BQCT scheme under PDN as below:

$$F^P = (1 - \eta_P)^7 + \eta_P^7 \alpha_0^4 \beta_{00}^4 + \eta_P^7 \alpha_1^4 \beta_{11}^4 \tag{26}$$

From (25) and (26), the fidelities for each of the two cases show that this is only depend on the amplitude parameter of the IS and the DR. Specifically, Figs. 2 a)-c) (Figs. d)-f)) clearly shows the effect of amplitude-damping (phase-damping) noise on the fidelity $F^A$ ($F^P$) and variation of the fidelity with amplitude parameter of the IS and the DR $\eta_r$. the fidelity $F^A$ and $F^P$ always decrease with decoherence $\eta_A$ and $\eta_P$, respectively as can be seen easily(Figs. 2 a), b), d), e)).

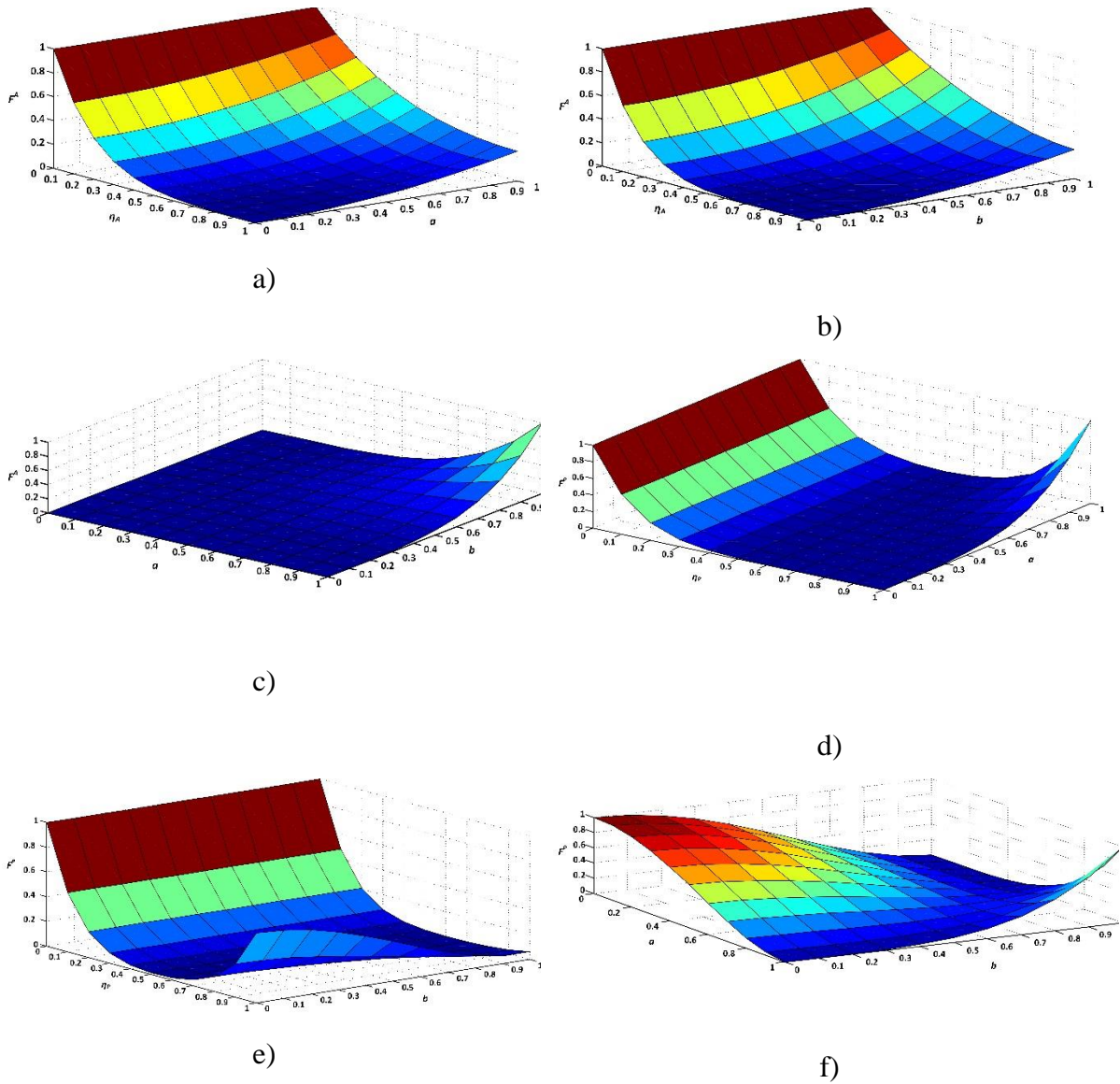

a)

b)

c)

d)

e)

f)

Fig. 2: Effect of noise on BQCT scheme is shown by fidelity FA variation (for amplitude damping noise model) and FP (for phase damping noise model) versus amplitude information of the states to be teleported and DRs for various situations: ADN with **a)** $a = \alpha_0, \beta_{00} = \beta_{11} = \frac{\sqrt{2}}{2}, \beta_{01} = \beta_{10} = 0$ **b)** $b = \beta_{00}, \alpha_0 = \alpha_1 = \frac{\sqrt{2}}{2}, \beta_{01} = \beta_{10} = 0$ **c)** $b = \beta_{00}, \eta_A = 1, a = \alpha_0, \beta_{01} = \beta_{10} = 0$, and PDN with **d)** $a = \alpha_0, \beta_{00} = 1, \beta_{01} = \beta_{10} = 0$, **e)** $b = \beta_{00}, \alpha_0 = \frac{1}{2}, \beta_{01} = \beta_{10} = 0, \alpha_1 = \frac{\sqrt{3}}{2}$. **f)** $a = \alpha_0, b = \beta_{00}, \beta_{01} = \beta_{10} = 0, \eta_P = 1$.

Also, it is observed a similar nature in Fig. 3, where the effect of ADN with PDN by assuming $\eta_A = \eta_P = \eta$ and $\alpha_0 = \alpha_1 = \frac{1}{\sqrt{2}}$, $\beta_{00} = \beta_{01} = \beta_{10} = \beta_{11} = \frac{1}{2}$ (Here $\alpha_0, \alpha_1, \beta_{00}, \beta_{01}, \beta_{10}, \beta_{11} \in R$) can be compared. In this situation, for the same value of DR $\eta$, the fidelity of amplitude damping channel (solid line in Fig. 3.a)) is always more than that of the phase-damping channel (dashed line in Fig. 3.a)). Therefore, as can be seen, for this particular choice of the amplitude parameter of the IS, loss of information is less when the travel QBs are transferred through the amplitude damping channel as compared to the phase-damping channel. However, this is not true generally, as shown in Fig. 3.b), where can be seen that for $\eta > 0.6708$ and $\alpha_0 = \beta_{00} = \frac{1}{2}, \alpha_1 = \beta_{11} = \frac{\sqrt{3}}{2}, \beta_{01} = \beta_{10} = 0$, the effect of phase-damping channel on fidelity is less than that of the amplitude-damping channel.

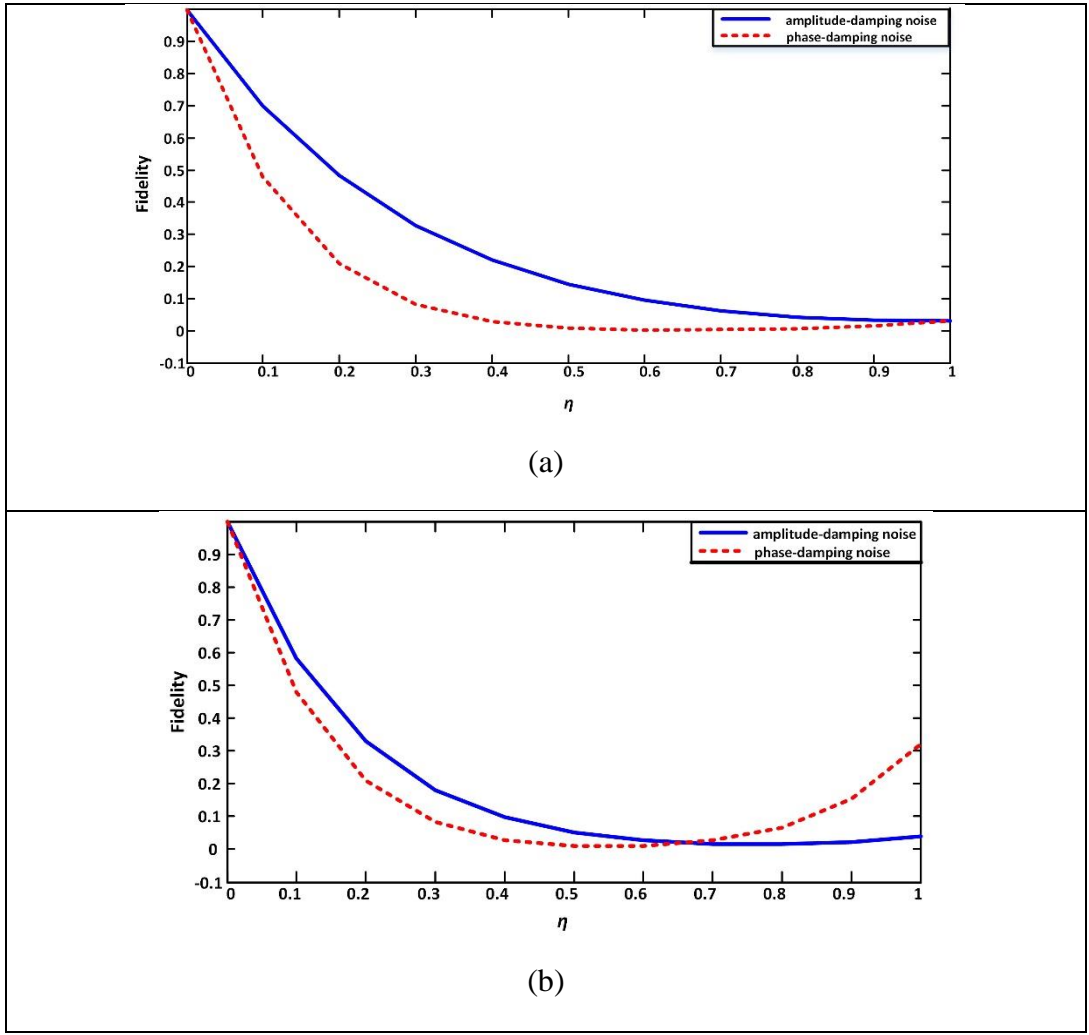

(a)

(b)

Fig. 3: The fidelities of ADN and PDN. The solid line stands for the ADN, the dashed line stands for the PDN

So, the fidelity decreases by increased noise in a real physical scenario. However, BQCT may be implemented with unit fidelity i.e. as prefect, if states with particular $\alpha_i$ and $\beta_{ij}$ are teleported, even in a NE. This fact shown in the peaks of Figs. 3 c), f).

**Comparison**

In this section, the proposed protocol is compared with the best presented BCQT protocols in terms of the type of protocol, the number of QBs sent by A and B, the number of QBs used in quantum channel, efficiency, BSMs (Bell-state measurements), SMs (single QB measurements), Prob. (i.e. the probability of guessing C's QB by an eavesdropper), and global operations (which for transmitting these gates, we need to add resources of entanglement states and QBs) as shown in Table 7. In this table, efficiency is defined [34] as the ratio of the number of sending QBs to the number of channel QBs. Also, in this table, QCPG stands for (Quantum Controlled Phase Gate).  As shown in Table 7, the proposed protocol only uses single QB measurement basis which is more efficient than two-QB measurements (Bell state measurements). It is well known that Bell-state measurements can be decomposed into an ordering combination of a SIQ Hadamard operation and a two-QB CNOT operation as well as two SIQ measurements. As shown in this Table, in this scheme, the users can teleport an arbitrary EN and a pure two-QBS to each other, simultaneously with the permission of a third party as supervisor or controller. As shown in this table, in some of the works [24, 27], the controller can only control one of the users. Also, in some of the works [25, 27], users need to apply global operations include of global CNOT gate and global QCPG between A and B. For transmitting these gates, as stated above, we need additional quantum and classical resources [35, 36]. So, these protocols are not optimal.

The work presented in [26] and the proposed protocol, both can teleport an EN and a pure two-QBS each to others simultaneously with the permission of a third party as supervisor or controller. However, the efficiency in our protocol is higher than [26]. Also, as stated above, the proposed protocol only uses eight SIQ measurements. But, work [26] used four two-QB measurements and one SIQ measurement. So, this work used a nine SIQ measurement that isn't optimal. In addition, our protocol reduces the probability of guessing C's QB by an eavesdropper to $\frac{1}{8}$. It is defined as the number of possible separate states obtained after the measurement by C as shown in Table 6. Also, the supervisor can control one of the users or both. In table 8, we compare the proposed protocol and protocol presented in [26] in resources used for preparing quantum channel in terms of the number of CNOT and Hadamard operations. As shown in this table, our protocol needs fewer resources (four to seven CNOT operations and three Hadamard operations) compared to the previous work [26] in the same conditions.

Table 7. Comparison of BCQT protocols

| Reference | Year | Type of protocol | B's QB | A's QB | Quantum channel | Efficiency | BSMs | SMs | Prob. | Global Operations |
|---|---|---|---|---|---|---|---|---|---|---|
| [24]* | 2016 | BCQT | 2 | 1 | 7 QB ENs | $\frac{3}{7}$ | 3 | 1 | $\frac{1}{2}$ | 0 |
| [25] | 2016 | BCQT | 2 | 1 | 7QB ENs | $\frac{3}{7}$ | 3 | 1 | $\frac{1}{2}$ | 1 CNOT |
| [26] | 2016 | BCQT | 2 | 2 | 9QB | $\frac{4}{9}$ | 4 | 1 | $\frac{1}{2}$ | 0 |
| [27]* | 2016 | BCQT | 1 | 2 | 6 QB cluster | $\frac{1}{2}$ | 2 | 2 | $\frac{1}{2}$ | 1 QCPG |
| Proposed Method | 2017 | BCQT | 2(EPR) | 2 | 8-QB ENs | $\frac{1}{2}$ | 0 | 8 | $\frac{1}{8}$ | 0 |

*In these protocols controller can only control one of users.

Table 8. Comparison of the prepared channel between the proposed method and [26].

| Reference | # CNOT operations | # Hadamard Operations |
|---|---|---|
| [26] | 12 | 5 |
| Proposed method | 4-7 | 3 |

## Conclusion

In this paper, a novel protocol was proposed for BCQT using of the eight-QB EN as the quantum channel by which the users can teleport an arbitrary EN and a pure two-QBS to each other simultaneously under the permission of the supervisor. This protocol was based on the controlled-not operation, appropriate SIQ UOs and SIQ measurements in the *Z* and *X*-basis which are more efficient than two-QB measurements [22, 26, 37]. In addition, in this protocol, the probability of guessing C's QB by eavesdropper was reduced and the supervisor can select control of one of the users or both. Also, in the proposed protocol, used quantum resources for preparing quantum channel and also, the number of measurements were fewer than previous works. Then, we reviewed the proposed protocol in typical noisy channels including the ADN and the PDNs. We analytically derived the fidelities of the BCQT process and showed that the fidelities only depend on the amplitude parameter of the IS and the DNR. We hope that such BCQT protocol can be realized experimentally in the future.

**Declarations of interest: none**